\documentclass[prd,eqsecnum,notitlepage,
nofootinbib,superscriptaddress]{revtex4-1}
\global\arraycolsep=2pt
\usepackage{stmaryrd}
\usepackage{amsmath}
\usepackage{amssymb}
\usepackage{graphicx}
\usepackage{textcomp}
\usepackage{calrsfs}
\usepackage{yfonts}
\usepackage{bm}
\usepackage{color}
\newcommand{\ben}{\begin{equation*}}
\newcommand{\een}{\end{equation*}}
\newcommand{\bean}{\begin{eqnarray*}}
\newcommand{\eean}{\end{eqnarray*}}

\newcommand{\be}{\begin{equation}}
\newcommand{\ee}{\end{equation}}
\newcommand{\bea}{\begin{eqnarray}}
\newcommand{\eea}{\end{eqnarray}}

\DeclareMathOperator{\tr}{tr}

\begin{document}
\title{Self-Stress  on a Dielectric Ball  and Casimir-Polder Forces}
\author{Kimball A. Milton}
  \email{kmilton@ou.edu}
  \affiliation{H. L. Dodge Department of Physics and Astronomy,
University of Oklahoma, Norman, OK 73019, USA}
\author{Prachi Parashar}
  \email{Prachi.Parashar@jalc.edu}
  \affiliation{John A. Logan College, Carterville, IL
62918, USA}
\affiliation{Department of Energy and Process Engineering,
Norwegian University of Science and Technology, 7491 Trondheim, Norway}

\author{Iver Brevik}
\email{iver.h.brevik@ntnu.no}
\affiliation{Department of Energy and Process Engineering,
Norwegian University of Science and Technology, 7491 Trondheim, Norway}


\author{Gerard Kennedy}\email{g.kennedy@soton.ac.uk}
\affiliation{School of Mathematical Sciences, 
University of Southampton, Southampton, SO17 1BJ, UK}

\begin{abstract}
It has always been conventionally understood that, in the dilute limit,
the Casimir energy of interaction between bodies or the Casimir
self-energy of a dielectric body could be identified  with the sum
of the van der Waals or Casimir-Polder energies of the constituents 
of the bodies. Recently, this proposition for self-energies
has been challenged by Avni
and Leonhardt [Ann.\ Phys.\ {\bf 395}, 326 (2018)], who find that the energy
or self-stress
of a homogeneous dielectric ball with permittivity $\varepsilon$ 
begins with  a term  of
order $\varepsilon-1$.  Here we demonstrate that this cannot be correct.
The only possible origin of a  term linear in $\varepsilon-1$ lies in the
bulk energy, that energy which would be present if either the material of
the body, or of its surroundings, filled all space.
 Since Avni and Leonhardt correctly subtract the bulk terms,
the linear term they find likely arises from their omission of an integral
over the transverse stress tensor.
\end{abstract}

\date\today
\maketitle

\section{Introduction}
\label{sec:intro}
From the beginning of the subject, it has been recognized that van der
Waals or Casimir-Polder forces between neutral atoms \cite{Casimir:1947hx}
and Casimir (also known as quantum-vacuum or dispersion) forces between
dielectric or conducting bodies
\cite{Casimir:1948dh} have the same origin.  This was made manifest
by Lifshitz \cite{Lifshitz:1956zz} and by Dzyaloshinskii, 
Lifshitz, and
Pitaevskii \cite{dlp} who generalized Casimir's calculation of
the force between perfectly conducting plates to that between dispersive
dielectric slabs.  In the dilute limit, the resulting energy is just the
sum of the Casimir-Polder energies.

Later it was shown that the Casimir self-energy of a dielectric ball
in the dilute limit  \cite{Brevik:1998zs} is identical to that obtained
by summing the van der Waals energies \cite{Milton:1997ky}.  Even though
the Casimir self-energy of a dielectric ball
of permittivity $\varepsilon$  is divergent, in the
dilute limit the first term, of order $(\varepsilon-1)^2$, is finite, and
unambiguously coincides with the sum of Casimir-Polder energies.
This quadratic dependence on the susceptibility reflects the nature of
the van der Waals two-body interaction, which is quadratic in the 
polarizability.

Thus it was surprising when Avni and Leonhardt \cite{al} re-examined
the Casimir self-energy for a homogeneous dielectric  ball
and found, after discarding divergent terms, that the Casimir energy 
 has a leading term of order $\varepsilon-1$ in the dilute limit. 
However, they did obtain
the correct result for a perfectly conducting spherical shell 
\cite{Boyer:1968uf},
or more generally, for the finite energy for a diaphanous or isorefractive
 ball (where the speed of light is the same inside and outside the ball)
\cite{Milton:2018hov}.  They
attribute the discrepancy to their different method of regularization: 
instead of point-splitting in time or transverse coordinates, they consider
the difference in the radial-radial stress tensor at
finite displacements inside and outside the spherical boundary.
However, we show that their method is erroneous. The only possible
source of a linear term is the bulk pressure, which they, like we, subtract.
In this paper we explicitly compute that bulk pressure, which indeed begins
with order $\varepsilon-1$, and show that it cannot give a finite remainder,
and that on physical grounds it must be subtracted.
What is left has to behave like $(\varepsilon-1)^2$ in the dilute limit. 
 Because of the error in
their method of computing the pressure on the  spherical surface, the
claim of Ref.~\cite{al} that the Casimir energy of the dielectric ball
``shutters the picture of the equivalence between the macroscopic effect
and pairwise summation'' is incorrect.  We can indeed
 ascribe the Casimir force to the sum of Casimir-Polder or van der Waals
energies, where in dense media pairwise summation must be supplemented by
multi-particle interactions.

\section{Bulk pressure on a dielectric ball}
\label{sec:ds}
We first examine the bulk pressure, as it is the only possible source
of a linear $\varepsilon-1$ term.
We consider a  ball, of radius $a$, 
made of homogeneous, isotropic  material, with permittivity 
$\varepsilon$ and permeability $\mu$ inside, surrounded by an infinite
background characterized by electrical properties $\varepsilon'$, $\mu'$.
The pressure on the spherical surface of the ball was first worked out in the
general case in Ref.~\cite{Milton:1996wm}:
\be
p=\frac1{2a^4}\int_{-\infty}^\infty \frac{dy}{2\pi} e^{iy\tilde\tau}
\sum_{l=1}^\infty \frac{2l+1}{4\pi}P_l(\cos\delta)
\left[x\frac{d}{dx} \ln D_l+f_l(x)-f_l(x')\right],\label{pressure}
\ee
where $\tilde\tau=\frac{\tau}a\to 0$ is a time-splitting regulator,
$\delta\to0$ is a transverse spatial (angular) regulator, 
\be
D_l=(s_l(x)e'_l(x')-s'_l(x)e_l(x'))^2-\xi^2 (s_l(x)e'_l(x')+s'_l(x)e_l(x'))^2,
\ee
with 
\be\xi=\frac{\sqrt{\frac{\varepsilon}{\varepsilon'}\frac{\mu'}{\mu}}-1}
{\sqrt{\frac{\varepsilon}{\varepsilon'}\frac{\mu'}{\mu}}+1},
\ee
and  $y=\zeta a$, $\zeta=-i\omega$ is the imaginary frequency,
 $x=|y|\sqrt{\varepsilon\mu}$, $x'=|y|\sqrt{\varepsilon'\mu'}$, while
the modified Riccati-Bessel functions are
\be
s_l(x)=\sqrt{\frac{\pi x}2}I_{l+1/2}(x),\quad
e_l(x)=\sqrt{\frac{2 x}\pi}K_{l+1/2}(x).
\ee
The terms subtracted in Eq.~(\ref{pressure}) are the terms present
 if either medium filled all of space:
\be
f_l(x)=2x[s'_l(x)e'_l(x)-e_l(x)s''_l(x)].
\ee
More precisely, this amounts to the removal of the pressure that the interior
medium would exert at $r=a$ if it filled all space, less the pressure that the
exterior medium would exert at $r=a$ if it filled all space.  In other words,
only the scattering part of the Green's function is included
\cite{Parashar:2018pds}.
We refer to the removal of the difference in the non-scattering
parts as the bulk subtraction.

The difference of bulk pressures
 vanishes if the speed of light both inside and outside the ball 
is the same, $\sqrt{\varepsilon\mu}=\sqrt{\varepsilon'\mu'}$, which is
true for a perfectly conducting shell of negligible thickness in vacuum,
or more generally, for a diaphanous or isorefractive ball.  We will discuss
this case in Sec.~\ref{sec:diaph}; we merely note here that Ref.~\cite{al}
agrees with the usual results in these cases. This agreement is beside
the point, because in these situations the pressure is an even 
function of $\xi$, so a
linear term can never arise.  But for a purely dielectric  ball 
the bulk terms play a crucial role.  We will here explicitly
evaluate those terms; in some sense, 
the failure to  exclude them properly is what gives rise to terms
proportional to $\varepsilon-1$ in Ref.~\cite{al}, as these are the only
possible source of such effects.

It is easy to evaluate these terms directly:
\be
p^{(0)}=-\frac1{16\pi^2a^4}\int_{-\infty}^\infty dy\,e^{iy\tilde \tau}
\sum_{l=1}^\infty(2l+1)P_l(\cos\delta)[f_l(x)-f_l(x')].\label{pzero}
\ee
Let's suppose the  ball
is characterized by $\varepsilon\ne  1$,
 $\mu=1$,
and is surrounded by vacuum. The term linear in $\varepsilon-1$ is
\be
p^{(0)}_1=-\frac{\varepsilon-1}{32\pi^2a^4}\int_{-\infty}^\infty dy\,
e^{iy\tilde \tau}
\sum_{l=1}^\infty(2l+1)P_l(\cos\delta)x\frac{d}{dx}f_l(x),\label{lineps}
\ee 
where we have now assumed that there is no dispersion,  and now
$x=|y|$. We use the summation formula
\be
\sum_{l=0}^\infty (2l+1)P_l(\cos\delta)e_l(x)s_l(y)=\frac{xy}\rho e^{-\rho},
\quad \rho=\sqrt{x^2+y^2-2 x y\cos\delta}.
\ee
This gives rather immediately
\be
p^{(0)}_1=-\frac{\varepsilon-1}{16\pi^2a^4}\int_0^\infty dx\,x\cos x\tilde\tau
\left(2+e^{-x\tilde\delta}\left[-2+x\left(\tilde\delta-4/\tilde\delta\right)
\right]\right),
\ee
where $ \tilde\delta=\sqrt{2}\sqrt{1-\cos\delta}\approx \delta$.
The term independent of $\tilde\delta$ is a Fresnel integral:
\be
\int_0^\infty dx\,x \cos x\tilde\tau=-\frac1{\tilde\tau^2},
\ee
and the remaining integrals are absolutely convergent:
\be
p^{(0)}_1=-\frac{\varepsilon-1}{8\pi^2 a^4}\left[-\frac1{\tilde{\tau}^2}
-\frac{{\tilde\delta}^2-\tilde{\tau}^2}{\left({\tilde\delta}^2+\tilde{\tau}^2
\right)^2}+
\frac{{\tilde\delta}^2-3\tilde\tau^2}{\left({\tilde\delta}^2+\tilde\tau^2
\right)^3}\left({\tilde\delta}^2-4\right)\right].
\ee
Evidently, spatial point-splitting is insufficient to regulate all divergences.
For simplicity, we now set $\delta=0$; then this reduces to
\be
p^{(0)}_1=-\frac{\varepsilon-1}{2\pi^2 a^4}\frac3{\tilde\tau^4},
\ee
which is precisely what would be obtained if $\delta$ were set equal to zero
from the outset.  Recalling that $\tilde\tau=\tau/a$, we see that this
corresponds to an energy given by
\be
4\pi a^2p^{(0)}_1=-\frac{\partial E_1^{(0)}}{\partial a}, \quad \mbox{or}\quad
E_1^{(0)}=\frac{\varepsilon-1}{\pi a}\frac2{\tilde{\tau}^4}.\label{esmalle}
\ee
Thus the portion of
the bulk term linear in $\varepsilon-1$  is purely divergent, with no
finite remainder, in contradistinction to the linear term
found in Ref.~\cite{al}.
To provide some kind of ``physical'' interpretation of the energy expression 
in Eq.~(\ref{esmalle}), one might note that the fundamental parameter in the 
energy expression is the ratio between the geometric size $a$ and the distance
$\tau$ ($=c\tau$) covered by light during the cutoff time $\tau$. 
Characteristic for a volume energy is that this parameter has to be
 raised to a power of four.

We can easily evaluate $p^{(0)}$ exactly from Eq.~(\ref{pzero}) in just the
same way:
\be
p^{(0)}=\frac1{4\pi^2a^4}\int_0^\infty dx\,x\left(\cos x\tilde\tau-\frac1
{\sqrt{\varepsilon}}\cos( x\tilde\tau/\sqrt{\varepsilon})\right)
\left[1+\frac{e^{-x\tilde\delta}}{x^2{\tilde\delta}^4}\left(2x\tilde\delta
\left(1+{\tilde\delta}^2/4\right)
+2x^2{\tilde\delta}^2\left(1-{\tilde\delta}^2/4\right)
\right)\right].\label{Pzero}
\ee
Doing the integrals as before, and  again
setting $\delta=0$ for simplicity, we have
\be
p^{(0)}=\frac1{\pi^2 a^4\tilde{\tau}^4}(1-\varepsilon^{3/2}).\label{pxe0}
\ee
Recalling that $\tau=a\tilde \tau$ is independent of $a$, this corresponds
to the energy
\be
E^{(0)}=-\frac43\frac1{\pi a\tilde\tau^4}(1-\varepsilon^{3/2}).\label{exe0}
\ee
When  expanded for small $\varepsilon-1$, this reduces to 
Eq.~(\ref{esmalle}).  The conclusion is that the bulk energy is purely
divergent.\footnote{The analogy 
is not true for the $\delta$-function
sphere \cite{Parashar:2017sgo}.  There the first-order term in the coupling
$\zeta_p$ is
\ben
E_1=\frac{\zeta_p}\pi\left(\frac1{\tilde\tau^2}+\frac{11}{24}\ln\tilde\tau
-0.345979\right).
\een
There is a finite part to this ``tadpole contribution,'' but it is not unique
because of the presence of the logarithmic divergence.
But even here, this first-order term is usually omitted as unphysical.}
Appendix \ref{appb} contains a generalization of this result.

It is straightforward to demonstrate explicitly that the bulk subtraction 
results in the absence from $p$ of any term linear in $\varepsilon-1$ in the 
dilute limit. Let
\begin{equation} 
p^*=\frac{1}{2a^4}\int_{-\infty}^{\infty}\frac{dy}{2\pi}e^{iy\tilde\tau}
\sum_{l=1}^{\infty}\frac{2l+1}{4\pi}P_l(\cos\delta) \,x\frac{d}{dx}\ln D_l
\end{equation}
be the pressure prior to the bulk subtraction, so that $p=p^*-p^{(0)}$, and 
let $p^*_1$ be the part of $p^*$ that is linear in $\varepsilon-1$ in the 
dilute limit. Noting that $D_l|_{\varepsilon=1}=1$ and $\left.
\frac{d}{d\varepsilon} D_l\right|_{\varepsilon=1}=-\frac12 f_l(x)$, where $x=|y|$, we immediately obtain
\begin{equation}
p^*_1= -\frac{\varepsilon-1}{32\pi^2 a^4} \int_{-\infty}^{\infty} dy\, 
e^{iy \tilde \tau} \sum_{l=1}^{\infty} (2l+1) P_l(\cos \delta)\, x 
\frac{d}{dx} f_l(x),
\end{equation}
which is identical to Eq.~(\ref{lineps}). 
Thus, $p^*_1-p^{(0)}_1=0$, and we can conclude 
that, as a consequence of the bulk subtraction, $p$ has no term linear in 
$\varepsilon -1$ in the dilute limit.

\section{Stress tensor and comparison with Ref.~\cite{al}}
Avni and Leonhardt \cite{al} base their calculation on the divergence
of the stress tensor.  Let us re-examine that here.  According to 
Ref.~\cite{Parashar:2018pds}, the time-averaged divergence of the stress
tensor is proportional to the inhomogeneity of the permittivity:
\be
{\overline{\bm{\nabla}\cdot \langle\mathbf{T}\rangle}}(\mathbf{r})=
\frac12\int_{-\infty}^\infty
\frac{d\zeta}{2\pi}\tr \bm{\Gamma}(\mathbf{r,r};\zeta)
\bm{\nabla}\bm{\varepsilon}(\mathbf{r},\zeta).\label{divst}
\ee
For an isotropic
 dispersionless homogeneous dielectric ball this gives the 
radial component of the force density
\be
f_r=-\overline{\bm{\nabla}\cdot \langle\mathbf{T}\rangle}(\mathbf{r})
\cdot\mathbf{\hat r}=
\frac{\varepsilon-1}{2}\int_{-\infty}^\infty \frac{d\zeta}{2\pi}
\tr\bm{\Gamma}(a,a;\zeta)\delta(r-a).
\ee
The volume integral of this gives the stress, which,  divided by the surface 
area, gives the pressure on the surface of the ball:
\be
p=\frac{\varepsilon-1}2\int_{-\infty}^\infty \frac{d\zeta}{2\pi}\tr \bm{
\Gamma}(a,a;\zeta).
\ee
This already has one factor of $\varepsilon-1$.
Now if we make the bulk subtraction, it is evident that the Green's 
dyadic vanishes if $\varepsilon=1$; there is no way a linear term could
arise.  This was shown explicitly already nearly 40 years ago
\cite{Milton:1979yx}; the explicit quadratic term was worked out
in Ref.~\cite{Brevik:1998zs}.  See also Refs.~\cite{lambiase,miltonbook}.

Reference \cite{al} computes the stress on the ball not by some
process of point-splitting but by computing the
 difference in the  radial stress  at
radii $a+\Delta$ and $a-\Delta$.
Power and logarithmic divergences appear as $\Delta\to0$.
The coefficients of these divergences are determined numerically, and
these divergent terms are then simply omitted, 
even though the logarithmic term can
modify the finite remainder, in an arbitrary way.  
That remainder is regarded as the finite
self-energy.  In the dilute limit, the divergent terms are proportional to 
$(\varepsilon-1)^2$, but,  surprisingly, 
the finite part starts with a linear
dependence, with the stress behaving approximately as $F=0.02 (\varepsilon-1)
\hbar c/a^2$.  How can this be?

In more detail, Avni and Leonhardt \cite{al} 
note that the radial component of the force density is not a total divergence:
\be
f_r=-(\bm{\nabla}\cdot \mathbf{T})\cdot\mathbf{\hat r}=-\bm{\nabla}\cdot
(\mathbf{T\cdot \hat r})+\frac1r(T_{\theta}{}^{\theta}+T_{\phi}{}^{\phi})
=\frac{F}{4\pi r^2}\delta(r-a),\label{divthm}
\ee
so if the stress tensor were finite at $r=a$, 
 the stress on the sphere  could be written as
\be
F=4\pi r_-^2 T_{rr}(r_-)-4\pi r_+^2 T_{rr}(r_+)+4\pi\int_{r_-}^{r_+}dr\,r\,
[T_{\theta}{}^{\theta}+T_{\phi}{}^{\phi}](r), \label{forcerpm}
\ee
where the integral is over a volume bounded by concentric spheres
drawn inside and outside the boundary of the dielectric ball, $r_-<a<r_+$.
(This result would also 
obtain by directly integrating Eq.~(15) of Ref.~\cite{al}.)
The stress tensor components inside and outside the boundary are finite,
divergent as $r_{\pm}\to a$,
\be
r_\pm^2 T_{rr}(r_\pm)=a^\pm_{-3}\Delta_\pm^{-3}+a^\pm_{-2}\Delta_\pm^{-2}
+a^\pm_{-1}\Delta_\pm^{-1}+b^\pm_0 \ln\Delta_\pm+a^\pm_0,
\ee
where $\Delta_\pm=|r_\pm-a|$.  Reference \cite{al} omits the divergent terms,
claiming the finite force  is
\be
F_m=4\pi (a^-_0-a^+_0),  
\ee
even though the presence of the logarithmic terms means that the finite part 
cannot be unique.
However, the volume integral over $T_{\theta}{}^{\theta}+T_{\phi}{}^{\phi}$ is 
mysteriously omitted.  In fact the latter is actually divergent; the integral
over $r$ in Eq.~(\ref{forcerpm}) does not exist. 
For example, for $\varepsilon-1$ small, 
\be
T_\theta{}^\theta+T_\phi{}^\phi\sim\mp\frac{\varepsilon-1}{160\pi^2(r-a)^4},
\label{angst}\ee
as the surface is approached from the outside (inside).  This singularity
is non-integrable.
(This is
discussed in detail in Appendix \ref{appa}.) Only if $\Delta^\pm\to0$,
and the integrals regulated in some other way, such as by a temporal 
point splitting as we do, can the integral over the angular components
of the stress tensor be omitted.\footnote{For example, using the leading
uniform asymptotic approximant given in Eq.~(\ref{angdiv}), we find the
integral in Eq.~(\ref{forcerpm}) goes like $ (r_+-r_-)/
(a^3\tilde\tau^4)\to 0$
as  $r_+-r_-\to 0$ for fixed $\tilde\tau$.
Similarly, there would be a divergent contribution
to the radial-radial integrated term in Eq.~(\ref{forcerpm}) as well, if regularization
were not supplied.} 
  Then Eq.~(\ref{forcerpm}) yields the
pressure (\ref{pressure}).  The arguments in Ref.~\cite{al}
about the invalidity of interchanging
the limits $r\to a$ and infinite series on $l$ are questionable.
The choice  of $\Delta_\pm$ as a putative regulator is a poor one;
$\Delta_\pm$ is ineffective in this capacity.

\section{Diaphanous or Isorefractive Ball}
\label{sec:diaph}
As mentioned above, in the case that the speed of light is the same inside
and outside the sphere, $\sqrt{\varepsilon\mu}=\sqrt{\varepsilon'\mu'}$, 
the bulk term vanishes.  So there is no question that the energy begins as 
$(\varepsilon-\varepsilon')^2$.  But Ref.~\cite{al} claims
small discrepancies with results shown in Ref.~\cite{Milton:1996wm}.  However,
the latter was just a first approximation, initially and more fully
 discussed by Brevik and Kolbenstvedt and other collaborators
\cite{Brevik:1982hc,brevik1982,Brevik:1984pt, brevik1987,
brevik1988, brevik1998}.  Recently, this simple problem has been reanalyzed 
with improved numerical results \cite{Milton:2018hov}.

In this case, the energy corresponding to the pressure (\ref{pressure}) reduces
 to 
\be
E=-\frac1{4\pi a}\int_{-\infty}^\infty dy\, e^{iy\tilde\tau}\sum_{l=1}^\infty
(2l+1)P_l(\cos\delta)x\frac{d}{dx}\ln(1-\xi^2[(e_l(x)s_l(x))']^2),
\ee
where now $\xi=(\varepsilon-\varepsilon')/(\varepsilon+\varepsilon')$,
and $x=|y|\sqrt{\varepsilon\mu}$. Reference \cite{al} notes 
that a $\Delta^{-1}$ divergence also occurs here; we saw the analog of
this previously for a general form of
cutoff in Ref.~\cite{Parashar:2017sgo}. However, for a purely temporal
regulator ($\delta=0$) the energy is completely finite.
 We can efficiently and accurately
evaluate this by including the first two uniform asymptotic approximants, 
followed by numerically computed remainders,
\be
E=\frac1{\sqrt{\varepsilon\mu}}\left(E^{(2)}+E^{(4)}+\sum_{l=1}^\infty R_l
\right),
\ee where the speed of light factor has been pulled out
(which Ref.~\cite{al} correctly notes we inadvertently omitted).  Here 
\be
E^{(2)}=\frac{5\xi^2}{32\pi a},\quad E^{(4)}=\frac{9\xi^2}{2^{12}a}\left(
\frac{\pi^2}2-1\right)(6-7\xi^2).
\ee
It is sufficient to include only the first two remainder terms, which gives
\be
E(\xi=0.33)=0.00380 (0.00387)/a; \quad E(\xi=0.2)= 0.001619 (0.00162)/a,
\ee
where the parenthetical numbers are those given in Ref.~\cite{al}.  There is
likely no significant discrepancy here.

\section{Discussion}
\label{sec:disc}
In this paper, we show that the procedure proposed in Ref.~\cite{al} is
erroneous.  One cannot extract the Casimir self-stress on the surface
of a homogeneous dielectric ball by taking the difference between the
normal-normal components of the stress tensor a finite distance away from
the boundary of the ball in the absence of regulation. 
We have demonstrated that although the stress tensor away from the
physical boundary
is finite,  divergences at the boundary and 
 a divergent integral over the angular parts of the stress tensor have
been rather mysteriously omitted.   Therefore the 
result of Ref.~\cite{al}, that there is a term linear in 
the susceptibilty in the dilute limit,  cannot be accepted.

Furthermore,
we have re-examined the issue of the bulk subtraction that must be supplied
in order to obtain a meaningful Casimir self-energy.  Doing so is necessary
to preserve the connection between van der Waals or Casimir-Polder forces and
the Casimir energy of a body composed of polarizable molecules. This is
signified by the energy in the dilute limit being proportional to 
$(\varepsilon-1)^2$.  A linear term can only arise if the bulk energy were not
properly subtracted.
We have evaluated the latter and have found that it only consists
of divergent terms; no meaningful linear term can be extracted.  Thus
there is no possible source of a term linear in the susceptibility. 

Unfortunately, the bulk subtraction is not sufficient to achieve a finite
Casimir self-energy for a  dielectric ball.  One way of stating this
fact is that the  $a_2$ heat-kernel coefficient is nonzero in
order $(\varepsilon-1)^3$ \cite{Bordag:1998vs}.  Only if the speed of light
is the same inside and outside a spherical shell of negligible thickness, 
which includes the case of the perfectly conducting spherical shell in
vacuum \cite{Boyer:1968uf}, can a unique, finite energy be extracted.  
Agreement with numerical results shown in 
Ref.~\cite{al} for such a case reveals no significant discrepancy with our 
results.

\acknowledgments
We thank our colleagues, Emilio Elizalde, Steve Fulling,
 Xin Guo, Yang Li, and  K. V. Shajesh,
for insightful discussions. We thank Yael Avni and Ulf Leonhardt
for comments prior to submission.  KAM acknowledges 
the financial support of the U.S. National Science Foundation,
grant No.~1707511 and PP and IB
 support from the Norwegian Research Council, project No.~250346. 

\appendix

\section{Scaling Behavior}\label{appb}

It is easily verified from Eqs.~(\ref{pressure}) and (\ref{pzero}) that both 
$p(\varepsilon, \mu, \varepsilon', \mu', a)$ and 
$p^{(0)}(\varepsilon, \mu, \varepsilon', \mu', a)$ possess the following 
scaling property, for scale factors $\lambda_{\varepsilon}, \lambda_{\mu} >0$:
\begin{equation}
p\left(\lambda_{\varepsilon}\varepsilon, \lambda_{\mu}\mu, 
\lambda_{\varepsilon}\varepsilon', \lambda_{\mu}\mu', (\lambda_{\varepsilon}
\lambda_{\mu})^{-\frac12} \,a\right)=(\lambda_{\varepsilon}
\lambda_{\mu})^{\frac32}\,p(\varepsilon, \mu, \varepsilon', \mu', a).\label{B1}
\end{equation}
This property can be useful. For example, it may be employed to relate the 
pressure for the dielectric-dielectric problem to that for the 
dielectric-vacuum problem: setting $\lambda_{\varepsilon}=
\frac{1}{\varepsilon'}$ and $\lambda_{\mu}=\frac{1}{\mu'}$ yields
\begin{equation}
p(\varepsilon, \mu, \varepsilon', \mu', a)=(\varepsilon'\mu')^{\frac32}\,p
\left(\frac{\varepsilon}{\varepsilon'}, \frac{\mu}{\mu'}, 1, 1, 
(\varepsilon'\mu')^{\frac12}\,a\right).
\end{equation}

The scaling property may be expressed in the form of a partial differential 
equation. To this end, it is helpful to work with an overall scale factor of
\begin{equation}
\lambda=\sqrt{\lambda_{\varepsilon}\lambda_{\mu}}
\end{equation}
and partition this between $\lambda_{\varepsilon}$ and $\lambda_{\mu}$ so that
\begin{equation}
\lambda_{\varepsilon}=\lambda^{\alpha} \quad \text{and} \quad \lambda_{\mu}=
\lambda^{\beta},
\end{equation}
where $\alpha+\beta=2$. Then Eq.~(\ref{B1}) becomes
\begin{equation}
p\left(\lambda^{\alpha} \varepsilon, \lambda^{\beta}\mu, \lambda^{\alpha}
\varepsilon', \lambda^{\beta} \mu', \lambda^{-1} a\right)=\lambda^3 \,
p(\varepsilon, \mu, \varepsilon', \mu', a). \label{B5}
\end{equation}
Differentiating with respect to $\lambda$ at $\lambda=1$, there results the 
following partial differential equation representation of the scaling property:
\begin{equation} 
a \frac{\partial}{\partial a}p-\alpha\left(\varepsilon\frac{\partial}{\partial 
\varepsilon}+\varepsilon' \frac{\partial}{\partial \varepsilon'}\right) p
-\beta\left(\mu\frac{\partial}{\partial \mu}+\mu' \frac{\partial}{\partial\mu'}
\right) p+3p=0,\label{B6}
\end{equation}
where $\alpha+\beta=2$. Note that
\begin{equation}
\left(\varepsilon\frac{\partial}{\partial \varepsilon}+\varepsilon' 
\frac{\partial}{\partial \varepsilon'}\right) p=\left(\mu
\frac{\partial}{\partial \mu}+\mu' \frac{\partial}{\partial \mu'}\right) p,
\end{equation}
because the functional dependence of $p$ on $\varepsilon, \mu, \varepsilon'$, 
and $\mu'$ is through only the following forms: $\varepsilon \mu$, $\varepsilon
' \mu'$, $\frac{\varepsilon}{\varepsilon'}$, and $\frac{\mu}{\mu'}$. 

The scaling property may also be expressed in the form of a conservation 
equation. Let 
\begin{equation} 
E= 4\pi a^3 p\left(\varepsilon, \mu, \varepsilon', \mu', a\right),\label{B8}
\end{equation}
which would be the energy if there were no cutoff dependence.
If, in Eq.~(\ref{B8}), $\varepsilon, \mu, \varepsilon'$, and $\mu'$ are 
permitted to vary with $a$ according to
\begin{subequations}
\begin{eqnarray} 
\varepsilon(a)&=&\varepsilon_1 \,a^{-\alpha},\\
\mu(a)&=&\mu_1 \,a^{-\beta},\\
\varepsilon'(a)&=&\varepsilon'_1\, a^{-\alpha},\\
\mu'(a)&=&\mu'_1 \,a^{-\beta},
\end{eqnarray}
\end{subequations}
where $\alpha+\beta=2$, then, as is easily verified from Eq.~(\ref{B6}), 
\begin{equation} 
\frac{d E}{da}=0,\label{B13}
\end{equation}
that is, $E$ is conserved under such change with $a$. (Here $\varepsilon_1,
\mu_1$, etc., are held constant, not the physical values $\varepsilon(a), 
\mu(a)$.)
It follows from Eqs.~(\ref{B8}) and (\ref{B13}) (or from Eq.~(\ref{B5}), with
$\lambda=a$) that
\begin{equation} 
p\left(\frac{\varepsilon_1}{a^{\alpha}}, \frac{\mu_1}{a^{\beta}}, 
\frac{\varepsilon'_1}{a^{\alpha}}, \frac{\mu'_1}{a^{\beta}}, a\right) = 
\frac{1}{a^3} \,p\left(\varepsilon_1, \mu_1, \varepsilon'_1, \mu'_1, 1\right).
\label{B14}
\end{equation}
Equation (\ref{B14}) provides a useful test of the functional form of terms in 
$p$. 

Let us illustrate this by examining the bulk pressure, which we write as
\begin{equation} 
p^{(0)}(\varepsilon, \mu, \varepsilon', \mu', a)=T_{rr}^{(0)}(\varepsilon, \mu,
 a_-)-T_{rr}^{(0)}(\varepsilon', \mu',a_+),\label{B15} 
\end{equation}
where (setting $\delta=0$)
\begin{equation} 
T_{rr}^{(0)}(\varepsilon, \mu, a)=-\frac{1}{2a^4} \int_{-\infty}^{\infty}
\frac{dy}{2\pi} \,e^{iy\tilde \tau}\sum_{l=1}^{\infty} \frac{2l+1}{4\pi} 
f_l(x).
\label{B16}
\end{equation}
$T_{rr}^{(0)}(\varepsilon, \mu, a)$ is the relevant component of the stress 
tensor in the case that the interior medium fills all space. It must therefore 
have dimensions of $\text{length}^{\!-4}$, yet be independent of $a$. This can 
only be achieved if it has a factor of $\tau^{-4}$. It is also a function of 
$\varepsilon$ and $\mu$ solely through the product $\varepsilon\mu$. In order 
to satisfy the scaling property (\ref{B14}), it must therefore have a factor 
of $(\varepsilon\mu)^{\frac32}$. These facts suggest that Eq.~(\ref{B16}) be 
rewritten as
\begin{equation} 
T_{rr}^{(0)}(\varepsilon\mu)=-\frac{(\varepsilon\mu)^{\frac32}}{\tau^4} 
g(\theta), \label{B17}
\end{equation}
where
\begin{equation} 
g(\theta)= \theta^4 \int_0^{\infty} \frac{dx}{2\pi}\,\cos(x\theta)\,
\sum_{l=1}^{\infty} \frac{2l+1}{4\pi} f_l(x) \label{B18}
\end{equation}
and
\begin{equation} 
\theta=\frac{\tau}{a\sqrt{\varepsilon \mu}}.\label{B19}
\end{equation}
By the above arguments, $g(\theta)$ must be independent of $a$. 
It must therefore be independent of its argument, that is, $g(\theta)$ must 
be a constant, $g$. To determine $g$, it suffices to apply Eq.~(\ref{B17}) in 
the case of vacuum, which immediately yields $g=\frac{1}{\pi^2}$.

This value can also be confirmed by explicit evaluation of Eq.~(\ref{B18}). If 
$\delta$ were reinstated, the method used in Eq.~(\ref{Pzero}) 
could be employed. However, here we will use an alternative approach. Let 
\begin{equation} 
f_l(x, r, s)=2\frac{\partial }{\partial r}\left(\frac{\partial}{\partial s}
-\frac{\partial}{\partial r}\right) \frac{s_l(rx) \,e_l(sx)}{x},\label{B20} 
\end{equation}
where $s>r>1$, which, by construction, satisfies
\begin{equation}
f_l(x)= \lim_{s\to r\to 1} f_l(x, r, s).\label{B21} 
\end{equation}
Then
\begin{equation}
\sum_{l=1}^{\infty}(2l+1) \,f_l(x, r, s)=2\frac{\partial}{\partial r}
\left(\frac{\partial}{\partial s}-\frac{\partial}{\partial r}\right) 
\left(\frac{rs}{s-r}-\frac{1}{2x}\right) e^{-(s-r)x}\label{B22}
\end{equation}
and
\begin{equation}
g=\frac{\theta^4}{8\pi^2} \lim_{s\to r\to 1} 2\frac{\partial }{\partial r}
\left[\frac{s-r}{(s-r)^2+\theta^2}+\left(\frac{\partial}{\partial s}
-\frac{\partial}{\partial r}\right)\frac{rs}{(s-r)^2+\theta^2}\right]=
\frac{1}{\pi^2}.\label{B23}
\end{equation}

Finally, it now follows from Eq.~(\ref{B15}) that
\begin{equation}
p^{(0)}=\frac{(\varepsilon'\mu')^{\frac32}-(\varepsilon\mu)^{\frac32}}
{\pi^2\tau^4},\label{B24}
\end{equation}
which generalizes Eq.~(\ref{pxe0}).

\section{Calculation of $T_\theta{}^\theta$}\label{appa}
Here we will examine the divergence structure of the transverse components of
the stress tensor as the surface of the sphere is approached.  In a medium
with constant $\varepsilon$ and $\mu$,
the stress tensor is defined by the dyadic
\be
\mathbf{T}=\bm{1}\frac12(\varepsilon E^2+\mu H^2)-(\varepsilon \mathbf{EE}
+\mu\mathbf{HH}), 
\ee
which yields for the transverse components
\be
T_\theta{}^\theta=\frac12\varepsilon(E_r^2+E_\phi^2-E_\theta^2)+\frac12\mu(
H_r^2+H_\phi^2-H_\theta^2)\to \frac12(\varepsilon E_r^2+\mu H_r^2),
\ee
where the last uses spherical symmetry.  Apart from a contact 
($\delta$-function) term, the divergenceless Green's dyadic $\bm{\Gamma}'
=\bm{\Gamma}+\bm{1}/\varepsilon$ can be constructed as follows:
\be
\bm{\Gamma}'(\mathbf{r,r'})=\sum_{lm}\left[\omega^2\mu F_l(r,r')\mathbf{X}_{lm}
(\Omega)\mathbf{X}_{lm}(\Omega')^*-\frac1\varepsilon \bm{\nabla}\times
G_l(r,r')\mathbf{X}_{lm}(\Omega)\mathbf{X}_{lm}(\Omega')^*\times\overleftarrow
{\bm{\nabla}'}\right],
\ee
where the vector spherical harmonics are defined by
$\mathbf{X}_{lm}=(l(l+1))^{-1/2}\mathbf{L}Y_{lm}$, $\mathbf{L}=\mathbf{r}
\times (1/i)\bm{\nabla}$.
Using the completeness property of the spherical harmonics
we immediately obtain
\be
T_\theta{}^\theta
=-\frac1{16\pi^2ar^2}\int_{-\infty}^\infty dy \,e^{iy\tilde\tau}
\sum_{l=1}^\infty (2l+1)l(l+1)P_l(\cos\delta)[G_l(r,r)+F_l(r,r)],\quad 
x=|y|\sqrt{\varepsilon\mu},
\label{ttt}
\ee
where we have retained the temporal and transverse spatial regulators $\tilde
\tau$ and $\delta$. This agrees with the formula (A.8) given in Ref.~\cite{al}.
Now we use the uniform asymptotic expansion as also given in Appendix B of
Ref.~\cite{al}.  The leading approximant gives close to the surface
of a homogeneous dielectric-diamagnetic ball in vacuum
($|r-a|\ll a$)
\be
F_l(r,r)\sim\pm\frac{R^E(z)}{2\nu r}t(z_\pm r/a)e^{-2\nu|r/a-1|/t(z_\pm)},
\ee
where the upper (lower) signs refer to the outside (inside) of the spherical
boundary, $\nu=l+1/2$, $z_+=z$, $z_-=z\sqrt{\varepsilon\mu}$, $t(z)=
(1+z^2)^{-1/2}$, and the reflection coefficient is
\be
R^E(z)=\frac{\sqrt{\mu z_-t(z_-)}-\sqrt{\varepsilon z_+t(z_+)}}
{\sqrt{\mu z_-t(z_-)}+\sqrt{\varepsilon z_+t(z_+)}}.
\ee  The TM Green's function $G_l$ is given by the same formula with $R^E\to
R^H$, obtained by interchanging $\varepsilon$ and $\mu$.

If we insert these approximants into Eq.~(\ref{ttt}) and omit the 
point-splitting parameters, we find outside or inside the sphere
\be
T^E_\theta{}^\theta (r)+T^H_\theta{}^\theta(r)
\sim \mp\frac3{64\pi^2}\frac1{(r-a)^4}
\int_0^\infty dz \frac{R^E(z)+R^H(z)}{(1+z_\pm^2)^{5/2}},\label{angdiv}
\ee
where the remaining integral on $z$ is evidently convergent.
So there is a quartic, nonintegrable singularity as $r\to a$.  Therefore, the
integral in Eq.~(\ref{forcerpm}) does not exist.  Note, for a perfectly
conducting sphere, $R^E\to -1$, $R^H\to +1$ so this leading singularity
cancels, and the leading surface divergence goes like $(r-a)^{-3}$ in agreement
with the well-known results 
\cite{Deutsch:1978sc,Kennedy:1979ar,Kennedy:1981yi}.

A similar calculation can be carried out for the radial-radial component of
the stress tensor.  In that case, the leading large-$\nu$ contribution cancels,
so the surface divergence is only of order $(r-a)^{-3}$ in general.  For
the perfect conductor case, again the TE and TM modes exhibit a further
cancellation, so the divergence is only of order $(r-a)^{-2}$.
[These results also follow from Eq.~(\ref{divthm}) for $r\ne a$.]  The energy
density and the transverse stress tensor components exhibit the most singular
behavior, as we would expect.

 The same types of cancellations,  $R^E+R^H=0$, occur for the 
isorefractive ball where $\varepsilon\mu=1$,  so $z_+=z_-$.
There, for example, the energy density behaves as \cite{Brevik:1984pt}
\be
u\sim \mp\frac{\mu-1}{\mu+1}\frac1{30\pi^2(r-a)^3},
\ee
as the surface is approached from the inside (outside).

\end{document}